\documentclass[11pt]{article}
\begin{document}
\def\a{\alpha}\def\b{\beta}\def\g{\gamma}\def\d{\delta}\def\e{\epsilon }
\def\k{\kappa}\def\l{\lambda}\def\L{\Lambda}\def\s{\sigma}\def\S{\Sigma}
\def\Th{\Theta}\def\th{\theta}\def\om{\omega}\def\Om{\Omega}\def\G{\Gamma}
\def\y{\vartheta}\def\m{\mu}\def\n{\nu}
\def\ws{worldsheet}
\def\susy{supersymmetry}
\def\ts{target superspace}
\def\ks{$\k$--symmetry}
\newcommand{\plabel}{\label}
\renewcommand\baselinestretch{1.5}
\newcommand{\nn}{\nonumber\\}\newcommand{\p}[1]{(\ref{#1})}
\renewcommand{\thefootnote}{\fnsymbol{footnote}}
\thispagestyle{empty}
\begin{flushright}
hep--th/0403100
\end{flushright}

\bigskip
\thispagestyle{empty}

\vspace{1cm}
\begin{center}
{\Large\bf On the sigma-model structure of type IIA supergravity
action in doubled field approach}

\vspace{2.5cm}
A. J. ~~Nurmagambetov
\footnote{Electronic address:  ajn@kipt.kharkov.ua}

\vspace{1.8cm}
{\small\it A.I. Akhiezer Institute for Theoretical Physics\\
NSC ``Kharkov Institute of Physics and Technology"\\
Kharkov, 61108, Ukraine}

\vspace{3.3cm} {\bf Abstract}
\end{center}
In this letter we describe how to string together the doubled
field approach by Cremmer, Julia, L\"u and Pope with
Pasti-Sorokin-Tonin technique to construct the sigma-model-like
action for type IIA supergravity. The relation of the results with
that of obtained in the context of searching for
Superstring/M-theory hidden symmetry group is discussed.

\vspace{3cm} {\it PACS: 04.65.+e, 04.50.+h, 11.25.Mj}

\renewcommand{\thefootnote}{\arabic{footnote}}
\setcounter{page}1

\newpage

The essential step towards making phenomenological sense of a
higher-dimensional theory is to find the relevant scheme of
reducing the extra dimensions and of a confining the Standard
Model fields on a four-dimensional space-time submanifold. The
idea of viewing this submanifold as a three-brane embedded into a
five-dimensional space \cite{rs} has been proved to be useful in
searching for the solution to some longstanding problems of
phenomenology like the hierarchy problem and the value of
cosmological constant. To shed a light on stringy origin of this
Brane-World picture \cite{bkorvp} one needs to deal with a domain
wall coupling to ten-dimensional bulk fields that enter the
low-energy effective action of a superstring theory. Constructing
such a coupling one should have in mind that as a superbrane the
domain wall is the source of antisymmetric tensor field of
supergravity multiplet and as the higher brane with the dimension
of worldvolume greater that five it couples to the conventional
supergravity tensor fields as well as to their duals. Moreover,
requirement of quantum consistency of a theory forces to take into
account other higher branes as a latent source of new local
anomalies. Hence coupling the higher branes to higher-dimensional
maximal supergravity backgrounds and studying the issue of anomaly
cancellations deserve of having a formulation for supergravities
whose dynamics is managed by the standard and the dual fields
entering the theory in a duality-symmetric way.

Another problem that is important from phenomenological point of
view is to identify the underlying symmetry group of
Superstring/M-theory in which the Standard Model group have to be
embedded. Recently it was emphasized that duality-symmetric
structure of maximal supergravities becomes important in searching
for such a hidden symmetry group
\cite{cj,julia,miz,west,h,kns,dhn,julia1}.

There are different routes to recover duality-symmetric structure.
For instance one can purely interested in dynamics of different
sub-sectors of duality-symmetric theory doubling the fields
on-shell, i.e. at the level of fields' equations of motion
\cite{cjlp}. After the doubling, applying a method akin to
non-linear realization technique one can find elegant
representation of the scalars' and antisymmetric tensors' duality
relations and equations of motion in the form of a twisted
self-duality condition and zero-curvature condition. Or one can
construct a supersymmetric pseudo-action \cite{bkorvp} supported
by on-shell duality relations between the doubled fields and these
relations have to be imposed by hands. There also is universal
formulation which is off-shell and produces the duality relations
as equations of motion \cite{bbs,dls,dlt,bns}. This formulation is
based on the ground of PST technique \cite{pst}. But either
pseudo-action approach or PST formalism which are just the
extensions of standard supergravity actions say nothing about the
hidden symmetry structure of maximal supergravities. On the
contrary, the doubled field approach of \cite{cjlp} is a
higher-dimensional remnant of $G/H$ coset structure of scalar
group manifolds that appear after toroidal reduction of D=11
supergravity with subsequent dualising the higher rank tensor
fields \cite{cjlp0}. In general the group of invariance of
equations of motion is more restrictive than the group of
invariance of the action. Therefore the step toward recovering the
true hidden symmetry group of Superstring/M-theory is the
construction of relevant low-energy effective actions. Since the
dynamics of scalars mentioned so far is described by $G/H$
sigma-model action one can also expect the sigma-model-like
structure of the doubled field action.

As a by-product of studying performed in \cite{bns} the doubled
field formalism of \cite{cjlp} dealing with non-gravitational
subsector of D=11 supergravity has been extended to the off-shell
supersymmetric formulation of duality-symmetric D=11 supergravity
\cite{bbs}. There has also been claimed that the sigma-model-like
structure of the doubled gauge field action in D=11 has a generic
form that remains the same also for the doubled field formulations
of type IIA and type IIB supergravities as well as for lower
dimensional maximal supergravities considered in \cite{cjlp}. This
point roughly seems to be clear at least for the sequence of
maximal supergravities coming from D=11 supergravity since all of
them are related to each other via dimensional reduction. However,
there is no rigorously defined procedure of dimensional reduction
for the sigma-model-like action. And having the duality-symmetric
gauge field structure in D=11 is not enough to recover complete
duality-symmetric structure of D=10 type IIA supergravity since
one shall double the fields coming from the reduction of gravity
sector. Hence the explicit supersymmetric sigma-model structure of
low-dimensional maximal supergravities  has to be shown in case by
case manner. The aim of the present paper is to fill the gap for
type IIA supergravity and to demonstrate how the generic form of
the D=11 doubled field sigma-model action \cite{bns} is
accommodated to describe its ten-dimensional counterpart.

Let us get started with recalling in brief  the doubled field
formulation of D=11 supergravity \cite{cjlp} and its extension at
the level of proper action \cite{bns}.
From the Lagrangian of D=11 supergravity \cite{cjs} (see
\cite{bns} for comprehensive list of our conventions)
$$
{\cal L}=-\sqrt{g}~R+{i\over 3!}{\bar\Psi}\wedge D[{1\over
2}(\omega+\tilde{\omega})]\Psi \Gamma^{abc}\wedge\Sigma_{abc}\\
-{1\over 48}\sqrt{g}~F_{mnpq}F^{mnpq} -{1\over 6}A^{(3)}\wedge
F^{(4)}\wedge F^{(4)}$$
\begin{equation}\label{eq3}
 -{1\over 2}(C^{(7)}+\ast C^{(4)})\wedge
(F^{(4)}+(F^{(4)}-C^{(4)})) ,
\end{equation}
with $F^{(4)}=dA^{(3)}$,
$C^{(4)}=-1/4~\bar{\Psi}\wedge\Gamma^{(2)}\wedge\Psi$,
$C^{(7)}=i/4~\bar{\Psi}\wedge\Gamma^{(5)}\wedge\Psi$,
$\Gamma^{(n)}=1/n!~dx^{m_n}\wedge\dots\wedge
dx^{m_1}\Gamma^{(n)}_{m_1\dots m_n}$, we get the second order
equation of motion for the $A^{(3)}$ gauge field
\begin{equation}\label{eq4}
d(\ast (F^{(4)}-C^{(4)})-{1\over 2}A^{(3)}\wedge
F^{(4)}-C^{(7)})=0
\end{equation}
that can be presented as the Bianchi identity for the dual field
$A^{(6)}$
\begin{equation}\label{eq5}
dF^{(7)}={1\over 2}F^{(4)}\wedge F^{(4)},\ \ \
F^{(7)}=dA^{(6)}+{1\over 2}A^{(3)}\wedge F^{(4)}.
\end{equation}

Now forget for a while the dynamical origin of
$F^{(7)}=dA^{(6)}+{1\over 2}A^{(3)}\wedge F^{(4)}$ and introduce
it as an independent partner of $F^{(4)}$. These field strengths
are invariant under the local gauge transformations
\begin{equation}\label{eq6}
\delta A^{(3)}=\Lambda^{(3)},\qquad \delta
A^{(6)}=\Lambda^{(6)}-{1\over 2}\Lambda^{(3)}\wedge A^{(3)}
\end{equation}
with closed forms $\Lambda^{(3)}$, $\Lambda^{(6)}$ associated with
the so-called large gauge transformations \cite{llps}. Because of
the presence of ``bare" $A^{(3)}$ in $F^{(7)}$ and therefore in
$\delta A^{(6)}$ that is traced back to the presence of the
Chern-Simons term in the Lagrangian (\ref{eq3}) the large gauge
transformations are non-abelian
\begin{equation}\label{eq7}
[\delta_{\Lambda^{(3)}_1},\delta_{\Lambda^{(3)}_2}]=\delta_{\Lambda^{(6)}},
\qquad [\delta_{\Lambda^{(3)}},\delta_{\Lambda^{(6)}}]=
[\delta_{\Lambda^{(6)}_1},\delta_{\Lambda^{(6)}_2}]=0.
\end{equation}
These relations can be associated with a superalgebra generated by
a ``Grassmann-odd" generator $t_3$ and a commuting generator $t_6$
\begin{equation}\label{eq8}
\{t_3,t_3\}=-t_6,\qquad [t_3,t_6]=[t_6,t_6]=0
\end{equation}
after that one can realize an element of the supergroup
\begin{equation}\label{eq9}
{\cal A}=e^{t_3 A^{(3)}} e^{t_6 A^{(6)}}
\end{equation}
and introduce the Cartan form
\begin{equation}\label{eq10}
{\cal G}=d{\cal A}{\cal A}^{-1}=F^{(4)}t_3+F^{(7)}t_6,
\end{equation}
which by definition satisfies the Maurer-Cartan equation called
sometimes the zero-curvature condition
\begin{equation}\label{eq11}
d{\cal G}+{\cal G}\wedge {\cal G}=0.
\end{equation}
To impose the duality relation between a priori independent field
strengths and arriving therefore at the standard number of degrees
of freedom one introduces the pseudo-involution operator ${\cal
S}$ which interchanges the generators $t_3$ and $t_6$
\begin{equation}\label{eq12}
{\cal S}t_3=t_6,\qquad {\cal S}t_6=t_3,\qquad {\cal S}^2=1.
\end{equation}
Using ${\cal S}$ and the Hodge operator one can immediately check
that the following condition
\begin{equation}\label{eq13}
\ast({\cal G}+{\cal C})={\cal S}({\cal G}+{\cal C}),
\end{equation}
where we have introduced the superalgebra valued element ${\cal
C}=-C^{(4)}t_{3}+C^{(7)}t_6$, reproduces correctly the duality
relations between the field strengths and therefore reduces
tensors' degrees of freedom to the correct number. Moreover, when
this condition holds the Maurer-Cartan equation amounts to second
order equations of motion for $F^{(4)}$ and $F^{(7)}$.

Applying the PST technique the twisted self-duality condition is
reproduced from the following action \cite{bns}
\begin{equation}\label{eq17}
S=S_{EH}+S_{\Psi}-Tr\int_{{\cal M}^{11}}\, [{1\over 4}\ast{\cal
G}\wedge {\cal G}+{1\over 2}({\cal G}+{1\over 2}{\cal C})\wedge
({\cal S}-\ast){\cal C}
\\-{1\over 12}{\cal G}\wedge {\cal S}{\cal
G}-{1\over 4}\ast i_v({\cal S}-\ast){\cal G}\wedge i_v({\cal
S}-\ast){\cal G} ],
\end{equation}
where $S_{EH}$ and $S_{\Psi}$ stand for the Einstein-Hilbert and
the Rarita-Schwinger actions,
\begin{equation}\label{eq18}
v={d a(x)\over \sqrt{-(\partial a)^2}}
\end{equation}
is the one-form constructed out the PST scalar auxiliary field
(cf. \cite{pst}) which ensures the covariance of the action and
\begin{equation}\label{eq19}
Tr(t_3 t_3)=-Tr (t_6 t_6)=-1,\qquad Tr(t_3 t_6)=0.
\end{equation}
As usual we have denoted by $i_v$ the inner product of the vector
field $v_m$ with a form.

It is an instructive exercise to rewrite the sigma-model action
(\ref{eq17}) to the standard for D=11 supergravity form. After
some manipulations with taking into account the definitions of
${\cal G}$ and ${\cal S}$ one can arrive at
\begin{equation}\label{eq21}
S=S_{CJS}+\int_{{\cal M}^{11}}\, {1\over 2}i_v{\cal
F}^{(4)}\wedge\ast i_v{\cal F}^{(4)}
\end{equation}
with $S_{CJS}$ being the standard action by Cremmer, Julia and
Scherk \cite{cjs} and ${\cal F}^{(4)}=(F^{(4)}-C^{(4)})-\ast
(F^{(7)}+C^{(7)})$. This is the action for the duality-symmetric
D=11 supergravity \cite{bbs} from which one can dynamically derive
the duality condition ${\cal F}^{(4)}=0$. Apparently, on the shell
of the duality condition ${\cal F}^{(4)}=0$ the action
(\ref{eq21}) coincides with the Cremmer-Julia-Scherk action.

Let us now turn to the type IIA supergravity which can be obtained
from D=11 supergravity by dimensional reduction. The corresponding
Lagrangian has the following form \cite{gp,cw,hn}
$$
{\cal L}=-\sqrt{-g}R-{i\over 3!}\bar{\psi}\wedge D[{1\over
2}(\omega+\tilde{\omega})]\psi \wedge
\Gamma^{abc}\Sigma_{abc}\\-{i\over 2}\bar{\lambda}\Gamma^a
D[{1\over 2}(\omega+\tilde{\omega})] \lambda \wedge \Sigma_a
$$
$$
+(-)^{n+1}\sum_{n=1}^{4}[{1\over 2}~e^{(-)^{n+1}\cdot \theta
(n-1)\cdot {(5-n)\over 2}\phi}F^{(n)}\wedge\ast F^{(n)}$$
$$
+(-)^n (C^{(10-n)}-e^{(-)^{n+1}\cdot \theta (n-1)\cdot {(5-n)\over
2}\phi}\ast C^{(n)})\times $$
$$
\wedge (F^{(n)}-{1\over 2}C^{(n)})]
$$
\begin{equation}\label{LIIA}
+{1\over 2}B^{(2)}\wedge dA^{(3)}\wedge dA^{(3)}+{\cal O}(f^4).
\end{equation}
Here we have denoted $F^{(1)}=d\phi$; $F^{(3)}=dB^{(2)}$ is the
field strength of the NS 2-form field $B^{(2)}$ and the RR field
strengths are defined as
\begin{equation}\label{bfF}
{\bf F}=d{\bf A}-F^{(3)}\wedge {\bf A}
\end{equation}
considering the formal sum ${\bf A}=\sum_{n=0}^3~A^{(2n+1)}$ and
extracting the relevant combinations. $\theta$ stands for usual
step function taking the one in the case of a positive argument.
The last term of (\ref{LIIA}) denotes the quartic fermion terms
which do not involve into the gravitino, dilaton, dilatino and
antisymmetric tensor gauge fields supercovariantization. The
explicit form of the fermion bilinears $C^{(n)}$ can be read off
\cite{bns}.

Extracting the dual field strengths by use of equations of motion
one arrives at
$$
F^{(7)}=dB^{(6)}-A^{(1)}F^{(6)}+{1\over 2}A^{(3)}dA^{(3)},
$$
\begin{equation}\label{IIAdfs}
F^{(9)}=dA^{(8)}-{3\over 4}F^{(8)}A^{(1)}+{1\over
2}B^{(2)}dB^{(6)}-{1\over 4}F^{(6)}A^{(3)},
\end{equation}
while the RR field strengths $F^{(6)}$ and $F^{(8)}$ are defined
as in (\ref{bfF}).

Instead of repeating the analysis of large gauge transformations,
constructing the associated superalgebra and introducing the
Cartan form via non-linear realization of the supergroup element
as it has been done before we will proceed further in a slightly
different way.

Following \cite{cjlp} close inspection of the Bianchi identities
for the dual field strengths shows that they can be obtained from
the zero curvature condition for the Cartan form  $${\cal
G}={1\over 2}d\phi\cdot t_0+\sum_{n=2}^4~e^{(-)^{n+1}{(5-n)\over
4}\phi}F^{(n)}\cdot
t_{n-1}$$$$-\sum_{n=6}^8~e^{(-)^{n+1}{(5-n)\over
4}\phi}F^{(n)}\cdot t_{n-1}-{1\over 2}F^{(9)}\cdot t_8,$$ with
taking into account the following superalgebra of generators
$$[t_0,t_1]={3\over 2}t_1,\qquad [t_0,t_2]=-t_2,\qquad
[t_0,t_3]={1\over 2}t_3,$$$$[t_0,t_5]=-{1\over 2}t_5,\qquad
[t_0,t_6]=t_6,\qquad [t_0,t_7]=-{3\over
2}t_7;$$$$[t_1,t_2]=-t_3,\qquad \{t_1,t_5\}=t_6,\qquad
[t_2,t_3]=t_5,$$$$[t_2,t_5]=-t_7,\qquad \{t_3,t_3\}=t_6,\qquad
\{t_1,t_7\}={3\over 8}t_8,$$
\begin{equation}\label{IIAsa}
[t_2,t_6]={1\over 4}t_8,\qquad \{t_3,t_5\}={1\over 8}t_8.
\end{equation}
Loosely speaking, the most part of the superalgebra can be
recovered by extraction of the relevant Bianchi identities for
dual field strengths from the zero-curvature condition
(\ref{eq11}). The rest of the algebra is restored from the graded
Jacobi identities. Since we have required the zero-curvature
condition, we can always present the Cartan form as ${\cal
G}=d{\cal A}{\cal A}^{-1}$ for the supergroup element ${\cal A}$,
the explicit expression for which can be read off \cite{cjlp}.

To present the type IIA supergravity action in a sigma-model-like
form one needs to write the former in the generating for such a
representation following expression
$$
S=S_{EH}+S_{\psi}+S_{\lambda}
$$
$$
+(-)^{n+1}\int_{{\cal M}^{10}}\, \sum_{n=1}^4 [{1\over
4}~e^{(-)^{n+1}\cdot \theta (n-1)\cdot {(5-n)\over 2}\cdot\phi}
F^{(n)}\wedge\ast F^{(n)}
$$
$$
+{(-)^n\over 2}(C^{(10-n)}-e^{(-)^{n+1}\cdot \theta (n-1)\cdot
{(5-n)\over 2}\cdot\phi} \ast C^{(n)})\times
$$
$$
\wedge (F^{(n)}-{1\over 2}C^{(n)})]
$$
$$
+\int_{{\cal M}^{10}}\, \sum_{n=6}^9[{1\over 4}~e^{(-)^{n+1}\cdot
\theta (9-n)\cdot {(5-n)\over 2}\cdot\phi} F^{(n)}\wedge\ast
F^{(n)}
$$
$$
+{(-)^{n+1}\over 2}(C^{(10-n)}-e^{(-)^{n+1}\cdot \theta
(9-n)\cdot {(5-n)\over 2}\cdot\phi}\ast C^{(n)})\times
$$
$$
\wedge(F^{(n)}+{1\over 2}C^{(n)})]
$$
$$
+(-)^{n+1}\int_{{\cal M}^{10}}\, {1\over
4}\sum_{n=1}^4~e^{(-)^{n+1}\cdot \theta (n-1)\cdot {(5-n)\over
2}\cdot\phi} i_v{\cal F}^{(n)}\wedge\ast i_v{\cal F}^{(n)}
$$
$$
-\int_{{\cal M}^{10}}\, {1\over 4}\sum_{n=6}^9~e^{(-)^{n+1}\cdot
\theta (9-n)\cdot {(5-n)\over 2}\cdot\phi} i_v{\cal
F}^{(n)}\wedge\ast i_v{\cal F}^{(n)}
$$
\begin{equation}\label{gfIIA}
+{1\over 2}\int_{{\cal M}^{10}}\, \sum_{n=1}^4~{1\over
3^{[{n+1\over 4}]}}~F^{(10-n)}\wedge F^{(n)}+{\cal O}(f^4),
\end{equation}
where $S_{EH}$, $S_{\psi}$, $S_{\lambda}$ stand for the kinetic
terms of graviton, gravitino and dilatino, and $[{n+1\over 4}]$
denotes the integer part of the number ${n+1\over 4}$. Here we
have denoted by ${\cal F}^{(n)}$ the duality relations between the
fields and their dual partners
\begin{equation}\label{IIAdr}
{\cal F}^{(n)}=\hat{F}^{(n)}+e^{(-)^n\cdot \theta (n-1)\cdot\theta
(9-n)\cdot {5-n\over 2}\cdot\phi}\ast \hat{F}^{10-n},\quad
n=1,\dots,4,6,\dots,9,
\end{equation}
with $\hat{F}^{(n)}=F^{(n)}-C^{(n)}$ for $n<5$ and ${\hat
F}^{n}=F^{(n)}+C^{(n)}$ for $n>5$.

Defining the traces between the same generators as follows
$$Tr(t_0t_0)=-Tr(t_8t_8)=-4,\ \  Tr(t_1t_1)=Tr(t_7t_7)=-1,$$
\begin{equation}\label{TrIIA}
Tr(t_2t_2)=-Tr(t_6t_6)=-1,\ \  Tr(t_3t_3)=Tr(t_5t_5)=-1,
\end{equation}
and setting other traces to zero it is matter to check that eq.
(\ref{gfIIA}) is presented as $$
S=S_{EH}+S_{\psi}+S_{\lambda}$$$$-Tr \int_{{\cal M}^{10}}\,[
(-)^{\theta (5-n)}\{ {1\over 4}~\ast {\cal G}\wedge {\cal G}
-{1\over 2}({\cal S}-\ast){\cal C}\wedge ({\cal G}+{1\over 2}{\cal
C})\}$$
$$-{1\over 4}\ast i_v ({\cal S}-\ast){\cal G}\wedge i_v ({\cal
S}-\ast){\cal G} ]$$
\begin{equation}\label{smIIA}
-{1\over 4} Tr \int_{{\cal M}^{10}}\, {1\over
3^{[{min~(n,k)+1\over 4}]}}~ (-)^n~{\cal G}\wedge {\cal S}{\cal
G}+{\cal O}(f^4).
\end{equation}
The presence of the coefficient with theta-function in front of
the second integral over a ten-dimensional manifold which after
evaluating the traces from (\ref{TrIIA}) becomes proportional in
particular to the sum of $F^{(n)}\wedge\ast F^{(n)}$ manages the
sign flips in the dilaton and gauge fields' kinetic terms, and the
coefficient in the last meaning term of (\ref{smIIA}) gives the
correct signs and coefficients to obtain the last meaning term of
(\ref{gfIIA}). To recover the structure of the supercovariant
terms for the $F^{(n)}$s we have introduced the superalgebra
valued element
$${\cal C}=-{1\over 2}C^{(1)}\cdot
t_0-\sum_{n=2}^4~e^{(-)^{n+1}{(5-n)\over 4}\phi}C^{(n)}\cdot
t_{n-1}$$$$-\sum_{n=6}^8~e^{(-)^{n+1}{(5-n)\over
4}\phi}C^{(n)}\cdot t_{n-1}-{1\over 2}C^{(9)}\cdot t_8.$$

The twisted self-duality condition
\begin{equation}\label{IIAtwsd}
\ast({\cal G}+{\cal C})={\cal S}({\cal G}+{\cal C})
\end{equation}
with the pseudo-involution ${\cal S}$ exchanging the generators
\begin{equation}\label{IIAS}
{\cal S} t_n=t_{8-n},\ \ \ n=0,\dots,3,5,\dots,8;\qquad {\cal
S}^2=1
\end{equation}
is reproduced from the action (\ref{smIIA}) as an equation of
motion. As soon as eq. (\ref{IIAtwsd}) holds the zero-curvature
condition encodes the second order equations of motion for fields
and their duals.

Therefore, we have found the explicit form of the sigma-model
representation of the type IIA duality-symmetric supergravity
action and have demonstrated that the form of this representation
obtained previously for the D=11 duality-symmetric supergravity
\cite{bns} is generic and needs just slight modifications to
accommodate the structure of type IIA supergravity. The main
difference between eleven and ten-dimensional cases is the
structure of quartic fermion terms. In the former case the terms
of this type can be absorbed into the corresponding supercovariant
quantities, while in the latter case there are additional quartic
fermion terms which do not involve into the supercovariantization
(cf. \cite{gp,cw,hn}).

In conclusion, let us discuss the relation of the results with
that of obtained in the context of searching for
Superstring/M-theory hidden symmetry group
\cite{julia,miz,west,h,kns,dhn,julia1}. It was realized long ago
\cite{cj} that the global symmetry groups of toroidally
compactified up to four space-time dimensions D=11 supergravity
fall into the class of exceptional groups $E_n$ with $n\le 7$.
Discovering the exceptional geometry of D=3 maximal supergravity
\cite{kns} gave one more evidence in favor of previously
conjectured \cite{julia1} (and Refs. therein), \cite{miz} $E_{10}$
hidden symmetry group of ``small tension limit" of M-theory
\cite{dhn} compactifying to one dimension. Recently it was
demonstrated that the bosonic sector of D=11 supergravity can be
reformulated as a non-linear realization of $E_{11}/F_{11}$ coset
space with maximal non-compact group $F_{11}$ containing
$SO(1,10)$ as a subgroup \cite{west}. To realize such a
formulation and to require representatives of such a coset space
belong to the Borel subgroup of $E_{11}$ one is forced to involve
dual partners for graviton and 3-index antisymmetric tensor gauge
field from the beginning and the action obtained is expected that
of a sigma-model. Additional strong arguments in favor of the
$E_{11}$ M-theory hidden symmetry group conjecture \cite{west}
follow from careful analysis of group structure relevant to
realize non-linearly the bosonic sector of type IIA supergravity.
The approach based on the non-linear realization is rigorous and
consistent but possesses the real drawback of absence of fermions
into the game. On the contrary, though the doubled gauge field
approach looks artificial it allows one to take fermions into
account. In common the abovementioned approaches seem to be
tightly related to each other, and the puzzle for both approaches
is to construct duality-symmetric action for graviton and its dual
partner. The necessity of making this step can be viewed for
instance under the derivation of duality-symmetric action for D=10
type IIA supergravity \cite{bns}. As it has been mentioned above
the Kaluza-Klein reduction of the duality-symmetric w.r.t. the 3-
and 6-index photons D=11 supergravity can not provide the
duality-symmetric structure of type IIA supergravity in the
subsector of fields coming from the reduction of metric tensor.
This fact forces to make additional efforts to recover complete
duality-symmetric formulation of D=10 type IIA supergravity. On
the other hand, having a formulation for D=11 supergravity which
is completely duality-symmetric w.r.t. the all fields including
the graviton, the complete duality-symmetric structure of D=10
type IIA theory could be recovered in a straight way. The
searching for the extension of a duality-symmetric formulation
with graviton field will shed a light on the relation between the
doubled field and non-linear realization approaches and will be
helpful in overcoming the drawbacks of the latter.

{\bf Acknowledgements.} We are very grateful to Igor Bandos and
Dmitri Sorokin for pleasant discussions and constant
encouragement. This work is supported in part by the Grant N
F7/336-2001 of the Ukrainian SFFR and by the INTAS Research
Project N 2000-254.

\end{document}